\documentclass[a4paper,12pt]{article}
\usepackage[dvips]{graphicx}
\usepackage{amsmath}
\usepackage{array}
\begin{document}
\begin{center}
{\large \bf The anisotropy properties of a background radiation in the fractal cosmological model}
\vspace{5ex}

I. K. Rozgacheva$^a$, A. A. Agapov$^b$
\vspace{2ex}

{\it VINITI RAS, Moscow, Russia

Peoples' Friendship University of Russia, Moscow, Russia

Moscow State Pedagogical University, Moscow, Russia}
\vspace{1ex}

E-mail: $^a$rozgacheva@yandex.ru, $^b$agapov.87@mail.ru
\vspace{4ex}
\end{center}
\begin{abstract}

We consider the anisotropy properties of a background radiation in the fractal cosmological model. The space of this model includes self-similar domains. The metric tensors of any two domains are connected by the discrete scaling transformation. Photons of the background radiation cross the domain and their energy change. Any observer receives these photons from different domains and detects spots with different brightness. The power spectrum of the brightness anisotropy of the background radiation in the fractal cosmological model is calculated. It is shown this spectrum is closed to the observed angular power spectrum of the SDSS-quasar distribution on the celestial sphere. Only qualitatively it conforms to the angular power spectrum of CMB (WMAP-7).
\end{abstract}
\vspace{1ex}

\vspace{3ex}

\begin{center}
{\bf 1. Introduction}
\end{center}
\vspace{1ex}

We call the fractal properties of the large-scale galaxy distribution and the CMB angular anisotropy by the fractal properties of the observable Universe. We have analyzed the fractal properties in our papers \cite{1}, \cite{2}, \cite{3} (see references there and the excellent review in book \cite{4}). We adduce here some power-laws indicating the fractality.

- The correlation dimension characterizes galaxy clumping degree and difference of the galaxy distribution from a homogenous and isotropic one. For example, the dependence of a SDSS-quasar number $N\left(r\right)$ in a sphere on its radius r is described by a power-law \cite{3}:
$$
N\left(\le r\right)\sim r^{d_c}, \eqno(1)
$$
where the exponent is the correlation dimension, it equals $d_c\approx 2{.}17$.

- The angular correlation function $\omega\left(\vartheta\right)$ and the angular power spectrum $u_l$ of the SDSS-quasar distribution approximate to power-laws at the average \cite{3}:
$$
\omega\left(\vartheta\right)\sim\vartheta^{-1{.}08}, \eqno(2)
$$
$$
u_l\sim l^{-1{.}08}, \eqno(3)
$$
where $l$ is a multipole moment number in expansion of the quasar distribution in spherical functions in a SDSS-survey part of the celestial sphere.

- Large-scale quasar clumps are discovered in the SDSS-quasar distribution. The relation between a number of clumps $N\left(\vartheta_c\right)$ and their angular size $\vartheta_c$ is characterized by a power-law \cite{3}:
$$
N_c\sim{\vartheta_c}^{-2{.}02}. \eqno(4)
$$

- The angular power spectrum of the CMB temperature fluctuations according to WMAP-7 data approximates to a power-law at the average \cite{3}:
$$
C_l\sim l^{-1{.}74}. \eqno(5)
$$

Power-laws (1) - (5) are considered to be indications of fractality because they may be interpreted through geometrical self-similarity of quasar clumps. For example, let angular sizes of clumps form a geometric progression
$$
\vartheta_i=q\vartheta_{i-1}, \eqno(6)
$$
where $q<1$ is a progression quotient. Then a number of quasar clumps with sizes $\vartheta_i\ge\vartheta_n=\vartheta_0q^{n-1}$ satisfies the dependence (4):
$$
N_n=\sum_{i=1}^nN_i\sim\sum_{i=1}^n\vartheta_i^{-d}=\vartheta_0^{-d}\frac{q^{-dn}-1}{q^{-d}-1}\approx\left(\vartheta_0q^{n-1}\right)^{-d}=\vartheta_n^{-d}.
$$

The correlation dimension value $d\approx 2{.}02$ for the size distribution of quasar clumps is compared to that of polygonal path of Brownian particle (length distribution of segments). This analogy indicates that, possibly, in quasar epoch the large-scale structure was composed of assembly of weakly interacting spatial regions. Quasar clumps discovered in our paper \cite{3} are physically self similar and they mark these spatial regions.

Sizes of any two regions are related by a scale transformation:
$$
r_i=qr_{i-1}. \eqno(7)
$$
Transformation (7) is an example of geometric self-similarity which is characteristic for mathematical fractals. Note that scale transformation (7) is discrete. In framework of field theory, this is a global transformation as opposed to local ones described by functions of spatial coordinates.

Discovered fractal properties (1) - (5) must be taken into account when formulating a theory of the large-scale structure formation in framework of the general relativity. They indicate that in this theory we must use solutions of Hilbert-Einstein equation which are invariant under discrete scale transformations of space-time metric.

The gravity theory in Riemannian spaces is known not to be invariant under local scale coordinate transformations of the Weyl group (conformal transformations) \cite{5} \cite{6}: local interval transformation
$$
ds^2\to \sigma\left(x\right)ds^2.
$$

In our paper \cite{7} the solution of Hilbert-Einstein equation and Lagrange's equation for a charged scalar meson matter field is constructed. The meson field in a field theory is described by a complex field $\psi$ with a rotary symmetry
$$
\psi\psi^*=\Psi^2=\mbox{const}, \eqno(8)
$$
where the asterisk denotes complex conjugation and $\Psi$ is the field amplitude related to the field charge $Q\sim\Psi^2$). In this case, Einstein's and Lagrange's equations are satisfied for the class of fields $\psi$ and $\tilde\psi$ of the form $\psi=\Psi\mbox{e}^{i\varphi}$ and $\tilde\psi=\tilde\Psi\mbox{e}^{i\tilde\varphi}$ which possess constant energy densities and related by the scale transformation (scaling):
$$
\Psi\leftrightarrow\frac1{\gamma}\tilde\Psi, \ \ \varphi\leftrightarrow\gamma\tilde\varphi, \eqno(9)
$$
where $\gamma$ is a numerical transformation parameter.

Field energy density and space-time metric $g_{mn}\left(\psi\right)$ allow the discrete transformation:
$$
E\leftrightarrow\frac1{\gamma^2}\frac{U_0}{\tilde U_0}\tilde E,
$$
$$
g_{mn}\left(\psi\right)\leftrightarrow \gamma^2\frac{\tilde U_0}{U_0}\tilde g_{mn}\left(\tilde \psi\right),
$$
where $U_0$ and $\tilde U_0$ are constant field potential parameters, $U=U_0\psi\psi^*$ and $\tilde U=\tilde U_0\tilde \psi\tilde \psi^*$. Therefore, space-time domains with field values related by the scaling (9) have geometrically similar metrics which differ in a constant factor only.

An obvious Newtonian analogy may be cited here. Let's consider two spherically symmetric bodies with masses $m_1$ and $m_2$. They produce gravity fields $\varphi_1\sim m_1\left.\right/ r$ and $\varphi_2\sim m_2\left.\right/ r$. Their spherical equipotential surfaces are geometrically similar and differ in radius of curvature only: the greater mass the less radius of curvature. Lengths of segments on the equipotential surfaces of these two fields are related by a discrete scaling. A transformation parameter of the scaling equals ratio of the bodies' masses.

Space-time geometry of the fractal cosmological model permits existence of fractal properties of the matter distribution which analogous to the observable fractal properties of the large-scale structure of the Universe (1) - (5).

Calculation of the power spectrum of the background radiation within framework of the fractal cosmological model \cite{7}, \cite{8} is performed in this paper.
\vspace{3ex}

\begin{center}
{\bf 2. Transfer of light signals}
\end{center}
\vspace{1ex}

A dynamic system of gravity and complex $\psi$ fields is described by Einstein-Hilbert action
$$
S=-\frac{c^3}{16\pi G}\int\left(R-\frac{8\pi G}{c^4}L\right)\sqrt{-g}\,d^4x,
$$
where $R$ is scalar curvature, $g<0$ is determinant of metric  tensor $g_{mn}$, space-time interval is $ds^2=g_{mn}dx^mdx^n$, indices take values 0, 1, 2, 3, metric signature is $(+---)$. We use the following form of the complex field Lagrangian:
$$
L=\frac1{hc}\left(g^{mn}\frac{\partial\psi}{\partial x^m}\frac{\partial\psi^*}{\partial x^n}-U\left(\psi\psi^*\right)\right),\eqno (10)
$$
where $U\left(\psi\right)$ is the field potential, $h$ is Planck's constant, $c$ is light velocity. Hereafter, the field dimension is $\left[\psi\right]=\mbox{erg}$, the contravariant metric tensor dimension is $\left[g^{mn}\right]=\mbox{cm}^{-2}$. This field possess the symmetry (8). Its Lagrange equation is
$$
\frac1{\sqrt{-g}}\frac{\partial}{\partial x^n}\left(\sqrt{-g}g^{mn}\frac{\partial\psi}{\partial x^m}\right)=-\frac{\partial U}{\partial\psi^*}.\eqno (11)
$$

In Einstein's equation
$$
R_n^m-\frac12R\delta_n^m=\kappa T_n^m \eqno(12)
$$
the energy-momentum tensor of the complex field equals to
$$
T_n^m=\frac{\partial\psi}{\partial x^n}\frac{\partial L}{\partial \left(\frac{\partial\psi}{\partial x^m}\right)}+\frac{\partial\psi^*}{\partial x^n}\frac{\partial L}{\partial \left(\frac{\partial\psi^*}{\partial x^m}\right)}-\delta_n^mL=\frac1{hc}g^{mp}\left(\frac{\partial\psi}{\partial x^p}\frac{\partial\psi^*}{\partial x^n}+\frac{\partial\psi}{\partial x^n}\frac{\partial\psi^*}{\partial x^p}\right)-\delta_n^mL,
$$
where $R_n^m$ is Ricci tensor, $\kappa =8\pi G\left.\right/c^4$ is Einstein's gravity constant, $G$ is Newton's gravity constant, $\delta_n^m$ is the Kronecker delta.

Following form of potential is used further:
$$
U=U_0\psi\psi^*.\eqno (13)
$$

One can ascertain through simple but rather cumbersome computations that Lagrange's equation (11) with potential (13) is satisfied for the solution:
$$
\psi =\Psi\mbox{e}^{i\varphi},\ \ \ \psi^*=\Psi\mbox{e}^{-i\varphi},
$$
$$
\Gamma_{mn}^l=\frac1{U_0}\frac{\partial^2\varphi}{\partial x^m\partial x^n}\left(g^{lp}\frac{\partial\varphi}{\partial x^p}+a^l\right),\eqno (14)
$$
$$
g_{mn}=\frac1{U_0}\left(4\frac{\partial\varphi}{\partial x^m}\frac{\partial\varphi}{\partial x^n}+\frac{\partial\varphi}{\partial x^m}a_n+\frac{\partial\varphi}{\partial x^n}a_m\right),
$$
where the field phase $\varphi\left(x^m\right)$ is a differentiable function. Hereafter, indices are raised and lowered with the metric tensor, indices appearing twice in a single term imply summing over its values, semicolon denotes covariant differentiation, $\Gamma_{mn}^l$ are Christoffel symbols.

Derivative $\displaystyle \frac{\partial\varphi}{\partial x^m}$ and covariant vector $a_m$ satisfy equations:
$$
g^{mn}\frac{\partial\varphi}{\partial x^m}\frac{\partial\varphi}{\partial x^n}=U_0, \ \ \ g^{mn}\left(\frac{\partial\varphi}{\partial x^m}\right)_{;n}=0, \eqno (15)
$$
$$
a_{m;l}=0,\ \ \ a_ma^m=-3U_0,\ \ \ \frac{\partial\varphi}{\partial x^m}a^m=0.
$$
Covariant $a_m$ and contravariant $a^k$ vectors satisfy equations:
$$
\frac{\partial a_m}{\partial x^l}=-3\frac{\partial^2\varphi}{\partial x^m\partial x^l},\ \ \ \frac{\partial a^n}{\partial x^m}a_n=3a^n\frac{\partial^2\varphi}{\partial x^n\partial x^m}.\eqno (16)
$$
One can ascertain through a substitution that the following equalities are satisfied for the solution (14) - (16):
$$
\frac{\partial g_{mn}}{\partial x^l}=g_{km}\Gamma_{nl}^k+g_{kn}\Gamma_{ml}^k,\ \ \ \Gamma_{kl}^m=\frac12g^{mn}\left(\frac{\partial g_{nk}}{\partial x^l}+\frac{\partial g_{nl}}{\partial x^k}-\frac{\partial g_{kl}}{\partial x^n}\right),\ \ \ \delta_m^n=g^{nl}g_{lm}.
$$
Functions $\displaystyle \frac{\partial\varphi}{\partial x^m}$, $a_m$, $a^m$ are derived from equations (11) and (16).

The solutions (14) are shown in paper \cite{7} to satisfy Hilbert-Einstein equation.

The Christoffel symbols and the Ricci tensor don't change under the discrete scaling (9). The $a_n$ vector, metric tensor, and components of the energy-momentum and Ricci tensors are multiplied by constant factors:
$$
a_n\leftrightarrow\gamma\tilde a_n,
$$
$$
g_{mn}\left(\psi\right)\leftrightarrow\gamma^2\frac{\tilde U_0}{U_0}\tilde g_{mn}\left(\tilde\psi\right),\eqno (17)
$$
$$
R_n^m\leftrightarrow\frac1{\gamma^2}\frac{U_0}{\tilde U_0}\tilde R_n^m,\ \ \ T_n^m\leftrightarrow\frac1{\gamma^2}\frac{U_0}{\tilde U_0}\tilde T_n^m.
$$
Therefore, Einstein's and Lagrange's equations are invariant under the discrete scaling (9).

The Lagrangian (10) equals to zero for the solution (14), whereas energy density is positive:
$$
E=\frac1{hc}\left(g^{mn}\frac{\partial\psi}{\partial x^m}\frac{\partial\psi^*}{\partial x^n}+U\left(\psi\psi^*\right)\right)=\frac2{hc}U_0\Psi^2>0.\eqno (18)
$$
The energy density (18) is constant, therefore the solution (14) corresponds to a stationary field condition. Space-time volumes with a structure of the solution (14) type are similar to each other and form a fractal set. Stationarity and fractality of solution (14) are consequences of the symmetry (8). Stationarity permits to refer this solution to the class of particle-like solution of the general relativity. Fractality implies that the solution corresponds to a set of noninteracting self-similar particles.

The phase path of the fields $\psi$ and $\psi^*$ is a circle.
$$
\psi\psi^*={\psi_1}^2+{\psi_2}^2=\Psi^2,
$$
$$
\psi=\psi_1+i\psi_2,\ \ \ \psi^*=\psi_1-i\psi_2.
$$
The function $\varphi$ is a degree of rotation round the circle. Length of a circle arc i.e. interval of set $\left\{\psi_1,\psi_2\right\}$ equals to
$$
dF^2=\left(d\psi_1\right)^2+\left(d\psi_2\right)^2=d\psi d\psi^*=\Psi^2\frac{\partial\varphi}{\partial x^m}\frac{\partial\varphi}{\partial x^n}dx^mdx^n. \eqno(19)
$$

Analyzing equations (14), (16) and formula (19) one can ascertain that there is a vector $a_m$ for which the metric $g_{mn}$ corresponds to a Riemannian space which is a tangent space of the phase path (8). This Riemannian space is an image of the phase space of the meson field and gravitational self-action of the meson field is a consequence of the symmetry (8).

Light signals transfer along isotropic geodesics of space-time (14). An isotropic vector satisfies equations:
$$
p_mp^m=0,\ \ \ p_{m;n}=0.
$$
Under solution (14) these equations are satisfied by the vector
$$
p_m=\sqrt{3}\frac{\partial\varphi}{\partial x^m}+a_m. \eqno(20)
$$
Under the discrete scaling (9) this vector transforms by the rule:
$$
p_m\leftrightarrow\gamma \tilde p_m. \eqno(21)
$$
Therefore, an isotropic vector remains isotropic under the scaling (9).
\vspace{3ex}

\begin{center}
{\bf 3. The power spectrum of a background radiation}
\end{center}
\vspace{1ex}

Self-similarity of space-time domains described by the solution (14) permits consideration of an assembly of such domains because transition from any domain to another resolves itself into dilatation or compression of an interval. Direction of isotropic geodesics doesn't change. We consider this assembly of self-similar space-time domains as the fractal structure of the cosmological model.


\begin{figure}[b!]
\begin{center}
\includegraphics[scale=1]{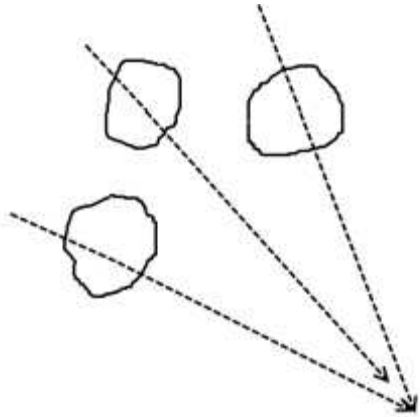}
\caption{\small Receiving of photons by an observer from different space domains}
\end{center}
\end{figure}

Let's consider transfer of photons of a background radiation within such fractal structure. We take into account gravitational influence of every domain only and leave out of account direct interaction with substance (absorption and scattering). In this case, energy of transferring photon changes due to gravitational frequency shift. When the photon is approaching a gravitating body its frequency is rising, when moving away the frequency is dropping.

Let the background radiation is homogeneous and isotropic, and an observer receives photons which passed through the fractal structure (fig.1). Brightness differs from one domain to another because energy of photons changes being multiplied by a scaling factor for each domain according to the expression (21). Therefore, the observer notices spots of different brightness in the distribution of the background radiation brightness on the celestial sphere.

Note that the relation of the CMB anisotropy with possible existence of spots of different brightness was discussed long ago \cite{9}, \cite{10}.

Compare the brightness distribution in the $j$-th spot with a distribution of points with angular distances to the spot center
$$
\vartheta=\arccos{\left(\cos{\delta_j}\cos{\delta}\cos{\left(\alpha-\alpha_j\right)}+\sin{\delta_j}\sin{\delta}\right)},
$$
where $\left(\delta,\alpha\right)$ and $\left(\delta_j,\alpha_j\right)$
are declinations and right ascensions of a point and of the spot center respectively. The whole spot is described by Legendre polynomial of order $l_j$ and its representation through spherical harmonics (the addition theorem):
$$
P_{l_j}\left(\cos{\vartheta_j}\right)=\frac{4\pi}{2l_j+1}\sum_{m=-l_j}^{m=+l_j}Y_{-m}^{l_j}\left(j\right)Y_m^{l_j}\left(\alpha,\delta\right). \eqno(22)
$$
The multipole number $l_j$ and the spot's size $\vartheta_j$ are related by the expression $
{\displaystyle \theta_j\approx\frac{\pi}{l_j}=\frac{180^{\circ}}{l_j}
}$. The polynomial (22) is of the bell shape in the range $0\le\vartheta_j\le\theta_j$ with maximum equal to 1 when $\vartheta_j=0$.

Summarized brightness distribution of the background radiation may be expressed now as
$$
F\left(\alpha,\delta\right)=\sum_{j=1}^N\gamma_jP_l\left(j,\alpha,\delta\right), \eqno(23)
$$
where the scaling factor $\gamma_j$ takes into account change of photons' energy when leaving the $j$-th domain. For determination of the power spectrum of the background radiation brightness distribution anisotropy the function (23) should be expanded in a spherical harmonics series:
$$
F\left(\alpha,\delta\right)=\sum_{m,l}a_{ml}Y_m^l\left(j,\alpha,\delta\right).
$$
The power spectrum is a function
$$
C_l=\frac1{2l+1}\sum_{m=-l}^{m=l}\left|a_{ml}\right|^2.
$$
Using definitions (22) and (23) we can determine the serial expansion coefficients $a_{ml}$:
$$
a_{ml}=\frac{4\pi}{2l+1}\sum_{j=1}^N\gamma_j\sum_{m=-l}^{m=l}Y_{-m}^l\left(j\right).
$$
The normalization condition for the spherical harmonics on a whole sphere is used here:
$$
\int Y_{-m}^{l_j}Y_m^ld\Omega=\delta_{ll_j}.
$$

In the simplest case of symmetric spots, the weight factors $\gamma_j$ are proportional to the spot's angular size, $\displaystyle \gamma_j\sim\theta_j\approx\frac{\pi}{l_j}$, and determine the dependence of the expansion coefficients on the multipole numbers: $\displaystyle a_{ml}\sim\sum_{j=1}^N\gamma_j\sim\sum_{j=1}^N\frac1{l_j}$. In this case, the power spectrum may be close to the power-law:
$$
C_l\sim\frac1{2l+1}\left(\sum_{j=1}^N\frac1{l_j}\right)^2\sim l^{-1}. \eqno(24)
$$

The model power spectrum (24) closely corresponds to the power spectrum of SDSS-quasars (3).
\vspace{3ex}

\begin{center}
{\bf 4. Conclusion}
\end{center}
\vspace{1ex}

The main result of the present work is computation of a background radiation power spectrum within the fractal cosmological model. The spectrum is shown to be close to the observable angular power spectrum of the SDSS-quasar distribution on the celestial sphere. It differs from the average power spectrum of the observable CMB anisotropy (WMAP-7). This fact will be a subject of investigation in further works.

The authors thank the FCPK grant 16.740.11.0465 for support of the work.

\end{document}